\documentclass[a4paper]{article}
\usepackage{INTERSPEECH_v2}

\usepackage[utf8]{inputenc}
\usepackage{amsmath,phonetic,url}
\usepackage{textcomp,algorithmic,epsfig,multirow,hhline}
\usepackage{amssymb,amsfonts}
\usepackage[table]{xcolor}

\title{Progressive Speech Enhancement with Residual Connections}

\name{Jorge Llombart, Dayana Ribas, Antonio Miguel, Luis Vicente, Alfonso Ortega, Eduardo Lleida}
\address{ViVoLab, Aragon Institute for Engineering Research (I3A) \\
University of Zaragoza, Spain}
\email{\{jllombg,dribas,amiguel,lvicente,ortega,lleida\}@unizar.es}

\begin{document}
\maketitle
%%%%%%%%%%%%%%%%%%%%%%%%%%%%%%%%%%%%%%%%%%%%%%%%%%%%%%%%%%%%%%%%%%%%%%%%%%%%%%%%%%%%%
\begin{abstract}
This paper studies the Speech Enhancement based on Deep Neural Networks.
The proposed architecture gradually follows the signal transformation during enhancement by means of a visualization probe at each network block. 
Alongside the process, the enhancement performance is visually inspected and evaluated in terms of regression cost. 
This progressive scheme is based on Residual Networks.
During the process, we investigate a residual connection with a constant number of channels, including internal state between blocks, and adding progressive supervision.
The insights provided by the interpretation of the network enhancement process leads us to design an improved architecture for the enhancement purpose. 
Following this strategy, we are able to obtain speech enhancement results beyond the state-of-the-art, achieving a favorable trade-off between dereverberation and the amount of spectral distortion.
\end{abstract}
%%%%%%%%%%%%%%%%%%%%%%%%%%%%%%%%%%%%%%%%%%%%%%%%%%%%%%%%%%%%%%%%%%%%%%%%%%%%%%%%%%%%%

%%%%%%%%%%%%%%%%%%%%%%%%%%%%%%%%%%%%%%%%%%%%%%%%%%%%%%%%%%%%%%%%%%%%%%%%%%%%%%%%%%%%%%
\section{Introduction}
%%%%%%%%%%%%%%%%%%%%%%%%%%%%%%%%%%%%%%%%%%%%%%%%%%%%%%%%%%%%%%%%%%%%%%%%%%%%%%%%%%%%%%
During the last few years, Speech Enhancement (\emph{SE}) based on Deep Neural Networks (\emph{DNN}) has emerged and positioned among the most active topics in the speech processing community.
Previous work evidences the ability of deep learning approaches for discovering underlying relations between the clean and the corrupted signal \cite{Xia2013,Feng2014,Tu2017,Karjol2018}. %, such that \emph{SE} solutions are able to accurately recover a clean version of the signal.
%
% However, one could wonder how the enhancement process is actually performed by the network. %(and if this inspection could guide us to influence in its performance) 
% %
% Maybe it is progressively ``cleaning'' the corrupted speech spectrum, or instead, it learns some special cues that induce it on a clean signal reconstruction.
%
The fact is that beyond what we know about the network main goal of minimizing the error between clean and corrupted signal, we actually are not sure on further ``why'' and ``how'' transformations are happening inside the network.
This black box effect is probably the major handicap/complaint against deep learning solutions, because it hinders the research process, and gives place to many empirical solutions.
Interpretability of deep neural networks has recently emerged as an area of machine learning research. 
It aims to provide a better understanding of how models perform feature selection and derive their classification decisions, such that the findings impact \emph{DNN} solutions design.
Recently, the top scientific conferences in the field have dedicated special spaces to this aim \cite{IRASL,IS2018}, evidencing the interest of the R\&D community. 

This paper is motivated to contribute to the accuracy and interpretability of \emph{SE} solutions. 
We present a \emph{SE} architecture following the feature-mapping strategy, where the enhancement process can be followed step by step by means of a visualization probe at each network block. 
%
% This progressive enhancement regularize the network improving the performance.
%
The visualization of the partial enhancement in each step allows us to supervise the process and collect relevant details on how it is performed.
This information is useful to detect which steps are meaningful in the enhancement, and which others can be discarded during the evaluation. 
%
% Accordingly, we can make our architecture variable in terms of performance/computation ratio even when the network is trained.
Therefore, even when the network is already trained, we can select a different grade of enhancement according to the application.
This way we have obtained a proper trade-off between accuracy and computational effort. 

The architecture proposed is based on the recent and powerful topology of Residual Networks (\emph{RN}) \cite{He2016} using one-dimensional convolution layers.
We have been able to get close to a performance in the state-of-the-art while exposing the step-by-step network processing at the same time.
The high potential of convolutional networks and residual connections have previously improved the performances for the speech enhancement task \cite{Qian2017,Park2017}. 
Residual connections provide the signal with a linear shortcut, while the non-linear path enhances it in several steps by adding or subtracting corrections. 
This mechanism could automatically compensate the amount of processing with the level of distortion in the signal.
This has also been tested in different more complex architectures using recurrent-based topologies \cite{Santos2018} or adversarial networks \cite{Pascual2017}. 
Recent work has studied the effect of residual and highway connections in speech enhancement models contributing to a better understanding of such models \cite{Santos2018a}. 
Contributions of this paper complement previous work on \emph{SE} using deep learning and the interpretability of these solutions. 
We search a better understanding of speech enhancement process by closely following the transformations carried out by the architecture. 
This way we contribute to improving the accuracy of the \emph{SE} through a novel architecture based on \emph{RN}.   
In the following, section \ref{sec:arch} presents the proposed architecture and its evolution through the study.
%
%Also, a preview of the qualitative results through the visualization of each network block in some signal examples.
%
Section \ref{sec:exp} describes the experimental setup designed for testing the \emph{SE} performance in reverberated speech through speech quality measures. 
Section \ref{sec:res} discusses results and finally section \ref{sec:conc} concludes the paper.

%%%%%%%%%%%%%%%%%%%%%%%%%%%%%%%%%%%%%%%%%%%%%%%%%%%%%%%%%%%%%%%%%%%%%%%%%%%%%%%%%%%%%
\section{Proposed architectures}
\label{sec:arch}
%%%%%%%%%%%%%%%%%%%%%%%%%%%%%%%%%%%%%%%%%%%%%%%%%%%%%%%%%%%%%%%%%%%%%%%%%%%%%%%%%%%%%

%-------------------------------------------------------------------------------------
\subsection{Constant Channel Residual Network}
\label{sec:wrn}
%-------------------------------------------------------------------------------------

\begin{figure}[h!]
%\centerline{\epsfig{figure=wrn.pdf,width=50mm}}
\centerline{\includegraphics[width=0.70\linewidth]{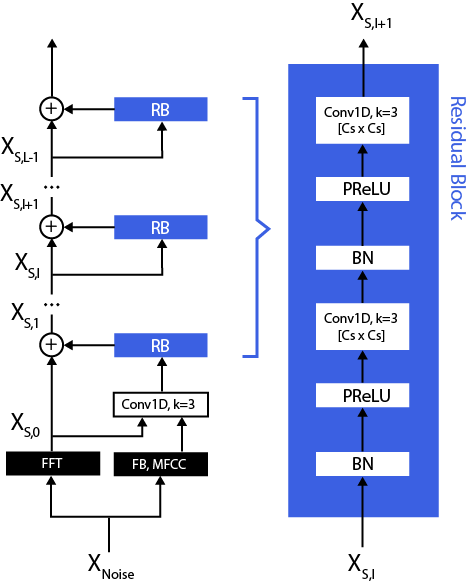}}
\caption{\footnotesize{\it Constant Channel Residual Network (\emph{CCRN}) architecture for progressive speech enhancement. $L=14$, $C_S=512$}}
\label{fig:wrn}
\end{figure}

In order to progressively enhance the input without losing the spectral representation in each block, we have designed an \emph{RN} that maintains the same number of channels throughout the residual connection. 
We will call this architecture Constant Channel Residual Network (\emph{CCRN}).
Figure \ref{fig:wrn} shows our system which uses multiple input sources to provide a variety of signal representations.
We provide multiple representations of the input in order to maintain as much reverberant impulse response inside the preprocessing analysis window as possible, without losing the temporal resolution of the acoustic events.
The front-end starts segmenting speech signals in 25, 50, and 75 ms Hamming window frames, every 10 ms. 
For each frame segment, three types of acoustic feature vectors are computed and stacked depending on the window width, to create a single input feature vector for the network: the logarithm of a 512-dimensional FFT, and 32, 50, 100-dimensional Mel filterbank \& cepstral features.
So the input feature vector has 876 dimensions.
Finally, each feature vector is variance normalized. 

The network processes input features with a first convolutional layer followed by $L=14$ Residual Blocks (\emph{RB}). 
This first layer uses the input dimension as the number of input channels, and the dimension of the logarithmic spectrum as the number of output channels. 
The \emph{RB} has two stages composed by a Batch Normalization (BN) layer \cite{Ioffe15}, a non-linearity by means of a Parametric Rectified Linear Unit (PReLU) \cite{He2015}, and a 1-dimension convolutional layer with a kernel of $k=3$ with the same number of channels at the output as at the input. 
The combination of \emph{BN} and \emph{PReLU} provides a smoother representation for regression tasks than the typical \emph{ReLU}.
% 
% This mechanism, instead of making a hard decision about the activation of the batch elements smaller than the batch mean, gets a trainable activation, that limits their impact in the result without dismissing their influence.
%
The residual connection is the addition of the input of the \emph{RB} and its output.
Our goal is to estimate the logarithmic spectrum of the clean signal (i.e. the enhanced $X_{S,L}$) from the logarithmic spectrum of the noisy signal.
Based on the experience from previous work \cite{llombart2018wide}, we process the full input signal as a sequence, instead of frame by frame.
We use a loss function based on Mean Square Error (\emph{MSE}) among frames (equation: (\ref{eq:jm2m})):
\begin{equation}
    J(Y, X_{S,L}) =\frac{1}{N}\sum^{N-1}_{n=0}\frac{1}{T}\sum^{T-1}_{t=0}MSE(y_{n,t}, x_{S,L,n,t})
 \label{eq:jm2m}
\end{equation}
\noindent where $N$ is the feature dimension, $T$ is the sequence length, $y_{n,t}$ are $Y$ frames and $x_{S,L,n,t}$ are $X_{S,L}$ frames.

\begin{figure}[h!]
%\centerline{\epsfig{figure=c31c204_ResNetCleanPath_column.pdf,width=70mm}}
\centerline{\includegraphics[width=0.9\linewidth]{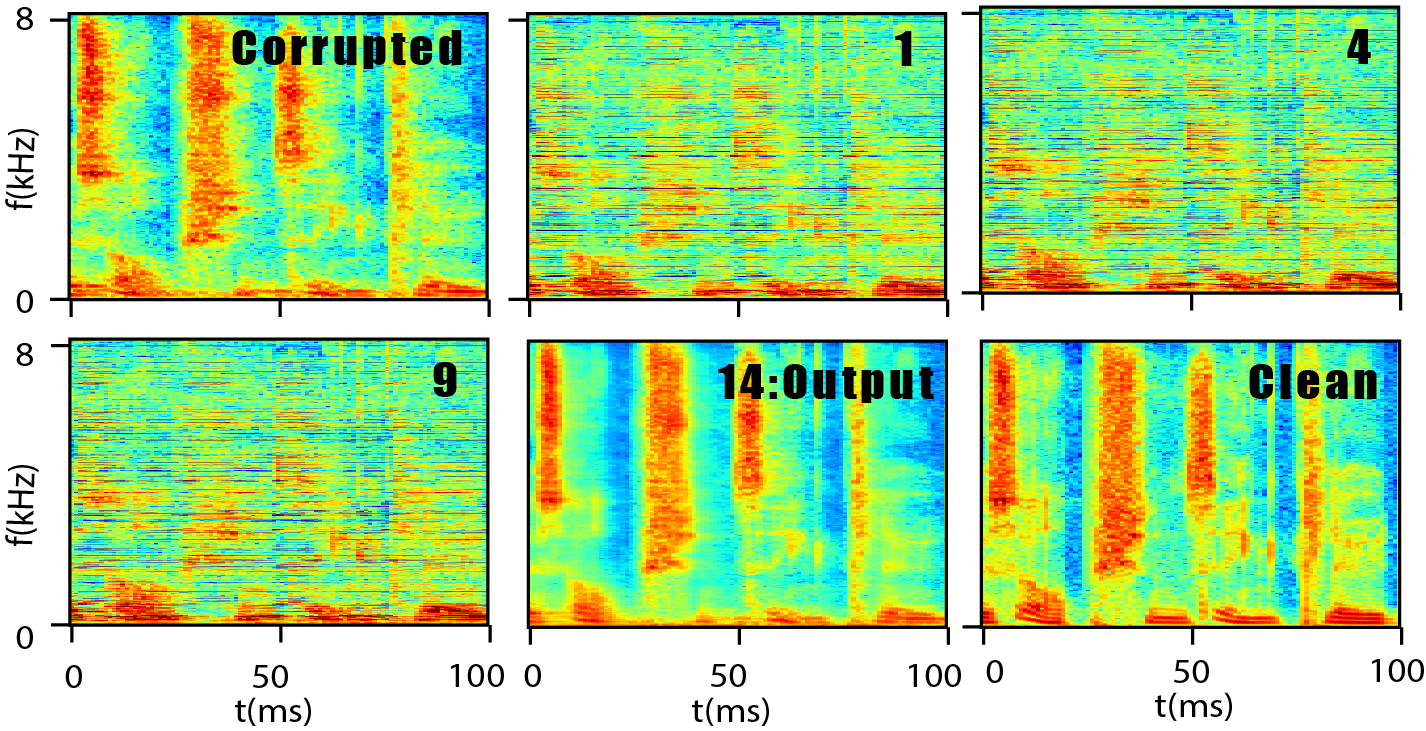}}
\caption{{\it Speech enhancement reconstructed output of progressive steps with \emph{CCRN} in a signal example.}}
\label{fig:wrnresult}
\end{figure}

%-------------------------------------------------------------------------------------
%\subsection{Layer-wise residual connection}
%\label{sec:wrn}
%-------------------------------------------------------------------------------------
Based on the previous description, we place a probe-output at each block, such that we can inspect the evolution of the enhancement process.
This is possible because we maintain unaltered the number of channels for each \emph{RB}.
Figure \ref{fig:wrnresult} shows an example of some steps output spectrum obtained.

%We can see that earlier blocks exhibit less frequency selectivity than later blocks, with the last block specializing strongly in certain frequencies. From block to block, we also see how the strongest components of the signal take shape, starting at the lower frequencies, followed by mid-bands and finally higher frequencies in the last stage

%
Note that the standard convolution layer is a linear combination of all input channels with a context for each output channel.
Therefore the resulting 512-dimensional matrix does not correspond to a proper spectrum. 
Also see that some frequency channels seem to group great part of the spectral information, while the rest seems to get blurred. 
This indicates the network focuses on certain frequencies channels consistently with the findings in \cite{Santos2018}.
This means the network steps over the spectral time-frequency structure, and only considers the weight changes from the \emph{MSE} minimization. 
This could be related to the different level of distortion among the frequency channels. 
Furthermore, see that the first and last step of the network processing is quite remarkable. 
The first step transforms the signal spectrum to strong or blurred frequency channels from the convolution output. 
While the last step suddenly reorders and recovers the signal with enhancement included. 
%
%-------------------------------------------------------------------------------------
\subsection{Constant Channel Residual Network with State Path} 
\label{sec:wrnctrl}
\begin{figure}[h!]
%\centerline{\epsfig{figure=wrn_ctrl.pdf,width=60mm}}
\centerline{\includegraphics[width=0.80\linewidth]{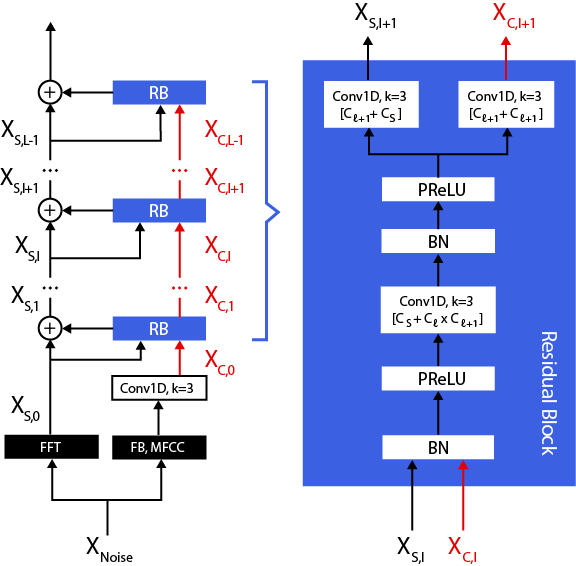}}
\caption{\footnotesize{\it Constant Channel Residual Wide Network with State path (\emph{CCRN-State}) architecture for progressive speech enhancement. $L=14$, $C_S=512$, $C_l=32*l, l \in [1, L]$ }}
\label{fig:wrnctrl}
\end{figure}

%-------------------------------------------------------------------------------------
The \emph{CCRN} architecture was designed to progressively enhance the input signal without losing the log-spectral representation.
However, despite we provide a shortcut path to allow the input to pass through with additive modifications, the training of the \emph{CCRN} makes a disorganized spectral representation. 
As we show in section \ref{sec:wrn} apparently most part of the information is grouped in some channels.
In order to pass the input over the residual path without changing its representation, we add a state path between \emph{RBs}.
In this way, the representation of the signal created by the network has its own path to going on. 
Moreover, this state path allows having more channels at each layer, while maintaining the same number of channels in the residual path.
We call the new architecture in figure \ref{fig:wrnctrl} Constant Channel Residual Network with State path (\emph{CCRN-State}).

This architecture stacks the channels of both paths at the input of the block. 
Inspired on the Wide Residual Networks \cite{Zagoruyko16}, we increase the number of channels in the first convolution of the block. 
Then, in order to obtain the residual path and a new state path, we use two convolutional layers at the block output.
One reduces the inner number of channels to the dimension of the residual connection, allowing the same behavior as in the previous architecture.
While the other provides the state path to be used by the next block.
However, in general, qualitative results showed again a spectrum with disorganized frequency channels. A similar pattern than the observed in the previous section. 

%-------------------------------------------------------------------------------------
\subsection{Progressive Supervision}
\label{sec:wrnMSE}
\begin{figure}[h!]
%\centerline{\epsfig{figure=c31c204_ResNetCleanPathAllMSE_column.pdf,width=70mm}}
\centerline{\includegraphics[width=0.9\linewidth]{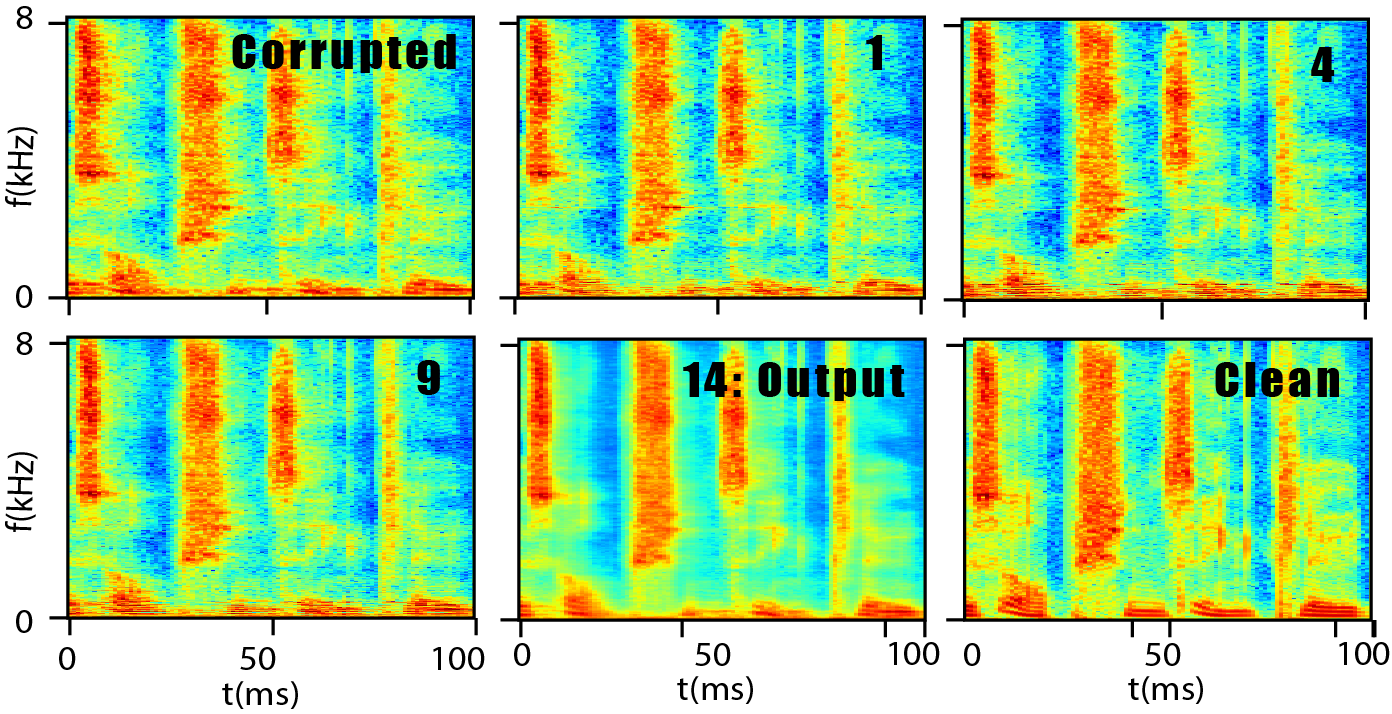}}
\caption{{\it Speech enhancement reconstructed output of different steps in a signal example forcing Progressive Supervision.}}
\label{fig:examplewithMSE}
\end{figure}

%--------------------------------------------------------------------------------------
Finally, to force the networks to provide a proper signal reconstruction at each step we add the \emph{MSE} cost term at each block output.
You can see antecedents of this strategy in classification tasks \cite{lee2015deeply}, although we are using it in a regression task.
In equation (\ref{eq:Jprogressive}), we add to the training cost the \emph{MSE} between the clean reference and each block output $X_{S,l} \forall l \in [1,L]$.
This second part in the cost function is weighted by $\alpha$. 
In our experiment, we choose $\alpha=0.1$ from development trials.
We call this cost \emph{Progressive Supervision}, because we take care of how much the network enhances the signal in each block.
\begin{equation}
    J_{PS}(Y, X_{S,L})=J(Y,X_{S,L}) + \alpha \frac{1}{L}\sum_{l=1}^L J(Y, X_{S,l})
    \label{eq:Jprogressive}
\end{equation}
Figure \ref{fig:examplewithMSE} shows a spectrogram example where we can see that finally we obtained an evolutionary pattern in the enhancement process. 
Note that both architectures explored in previous sections \ref{sec:wrn} and \ref{sec:wrnctrl} will have the same output representation because throughout the residual path there is a constant number of channels. 
Anyway, at the end of the processing, enhancement results could be different. 

We can see the earlier blocks take care of the more noticeable distorted areas of the spectrum, e.g. look at the trail of the reverberation. 
Apparently, the network is establishing a pattern of what is distortion and what is not. 
We also note that the network mainly focuses on the spectrum valleys and gradually the granularity in them is removed. 
Also, see how the spectral trail effect because of reverberation, is gradually removed. 
After last blocks, the network starts softening the spectrum in order to produce slow spectral magnitude changes. 
This avoids undesirable auto-generated distortions such as the annoying musical noise. 
However, it could also have an over-softening effect that causes an unrealistic effect in the final output (see block 14 output). 
%
% To compensate this, we could make the network learn about the concept of comfort noise, such that it could combine the \emph{MSE} optimization with a realistic spectral output. 

%

%
So far, the interpretative analysis of the enhancement process does not allow us to be totally sure about the impact on \emph{SE} performance.
In the following sections, we will assess the accuracy of the proposed model in an objective manner. 
Note that due to space issues the analysis so far used a single signal example. 
However, the observations discussed are common to different sentences and distortion types. See more examples in \url{https://medium.com/vivolab}. 
%https://medium.com/vivolab/progressive-speech-enhancement-with-residual-connections-e5f5d9735d89

%%%%%%%%%%%%%%%%%%%%%%%%%%%%%%%%%%%%%%%%%%%%%%%%%%%%%%%%%%%%%%%%%%%%%%%%%%%%%%%%%%%%%
\section{Experimental setup}
\label{sec:exp}
%%%%%%%%%%%%%%%%%%%%%%%%%%%%%%%%%%%%%%%%%%%%%%%%%%%%%%%%%%%%%%%%%%%%%%%%%%%%%%%%%%%%%
Speech enhancement system was developed using the Pytorch toolkit \cite{paszke2017automatic}.
Input examples for training were generated on-the-fly, distorting contiguous random sequences of 200 samples from Timit\cite{garofolo1993darpa}, Librispeech\cite{panayotov2015librispeech} and Tedlium\cite{rousseau2014enhancing} databases. 
AdamW algorithm was used to train the network \cite{Kingma2015,Loshchilov2017}. 
Approaches were tested on the official Development and Evaluation sets of the REVERB Challenge \cite{Challenge2013}.
The dataset has simulated speech from the convolution of WSJCAM0 Corpus \cite{WSJCAMO} with three measured Room Impulse Responses (\emph{RIR}) ($RT_{60} = 0.25, 0.5, 0.7 s$) at two speaker-microphone distances: far ($2 m$) and near ($0.5 m$).
It was added stationary noise recordings from the same rooms (SNR = 20 dB).
Besides, it has real recordings, acquired in a reverberant meeting room ($RT_{60} = 0.7 s$) at two speaker-microphone distances: far ($2.5 m$) and near ($1 m$)  from the MC-WSJ-AV corpus \cite{MC-WSJ-AV}. 
We also used real speech samples from VoiceHome v0.2 \cite{VoiceHome1} and v1.0 \cite{VoiceHome2}.
VoiceHome was recorded in a domestic environment from 3 real homes, such that the background noise is the one typically found in households e.g. dishwasher, vacuum or television.

%-------------------------------------------------------------------------------------
\subsection{Reference and performance assessment}
%-------------------------------------------------------------------------------------
We compare the performance with the state-of-the-art dereverberation method called Weighted Prediction Error (\emph{WPE}), which is known to effectively reduce reverberation and greatly boosts the speech enhancement performance. 
We used the more recent version of \emph{WPE} \cite{NaraWPE} which is also based on \emph{DNN} \cite{Kinoshita2017}. 
However, \emph{WPE} uses a different architecture based on \emph{LSTM}.
In order to assess the enhancement, we measure the distortion reduction through Log-likelihood ratio (also known as Itakuta distance) (\emph{LLR}) \cite{Loizou2011} computed in the active speech segments.
The closer the target feature to the reference, the lower the spectral distortion, therefore smaller values indicate better speech quality. 
On the other hand, we assess the reverberation level of the signal through Speech-to-Reverberation Modulation energy Ratio (\emph{SRMR}) \cite{SRMRmeasure}. 
In this case, higher values indicate better speech quality. 
%Note that only SRMR can be used with real data because the other quality measures are computed using the observed/enhanced signal and clean reference.

% \subsection{Network configuration}
% Speech enhancement system was built on the Pytorch toolkit \cite{paszke2017automatic}.
% %
% Input features for training the network were generated on-the-fly, operating in contiguous sequences of 200 samples, so that convolutions in the time axis can be performed. 
% %
% AdamW algorithm was used to train the network \cite{Kingma2015,Loshchilov2017}. 

%%%%%%%%%%%%%%%%%%%%%%%%%%%%%%%%%%%%%%%%%%%%%%%%%%%%%%%%%%%%%%%%%%%%%%%%%%%%%%%%%%%%%%
\section{Results and Discussion}
\label{sec:res}
%%%%%%%%%%%%%%%%%%%%%%%%%%%%%%%%%%%%%%%%%%%%%%%%%%%%%%%%%%%%%%%%%%%%%%%%%%%%%%%%%%%%%%
%-------------------------------------------------------------------------------------
\subsection{Speech quality for processing tasks}
%-------------------------------------------------------------------------------------
Table \ref{tabdistortion} presents speech quality results in terms of distortion. 
The first row corresponds to the reverberant unprocessed speech compared to the quality achieved using \emph{WPE} or \emph{CCRN}-based architectures.
All enhancement methods that appear in this paper were able to enhance the corrupted speech data, but our proposed architectures outperform \emph{WPE} in terms of distortion. 
In spite of the \emph{CCRN-State} based architectures have more degrees of freedom note that \emph{CCRN + Progressive Supervision} reache the best performance.  

\begin{table} [h]
\footnotesize
\caption{\footnotesize \label{tabdistortion} {\it LLR distance in simulated reverberated speech samples from REVERB Dev \& Eval datasets.}}
\centerline{
\begin{tabular}{c||c|c}
Methods & REV-Dev & REV-Eval \\  
\specialrule{.2em}{.1em}{.1em} 
Unprocessed & 0.63 & 0.64 \\ \hline       
WPE \cite{NaraWPE} & 0.60 & 0.60 \\ \hline 
CCRN & 0.52 & 0.53 \\ 
+Prog Sup & \cellcolor{gray!35}0.49 & \cellcolor{gray!35}0.49 \\ \hline 
CCRN-State & 0.51 & 0.53 \\   
+Prog Sup & 0.53 & 0.54 \\ 
\end{tabular}
}
\end{table}
%\vspace{-0.5cm}
%-------------------------------------------------------------------------------------
\subsection{Speech quality for dereverberation: Simulated vs. Real}
%-------------------------------------------------------------------------------------
%
Table \ref{resultsrealtest} shows the average of SRMR results over the evaluated conditions for simulated and real speech samples. 
The first row corresponds to the unprocessed speech data. 
Note that the best performances for all datasets evaluated were achieved by \emph{CCRN + Progressive Supervision}, consistent with the previous result for \emph{LLR}. 
The consistency in performance through different datasets supports the robustness of the method, indicating that its parameters are not adjusted to some specific set of speech signals. 
Furthermore, positive results beyond simulated reverberated speech encourage the use of this method in realistic scenarios. 
Moreover, note that all \emph{CCRN} models were trained with artificially synthesized reverberation, however, they showed to be effectively dealing with a reverberated speech from real-world scenarios.

\begin{table} [h]
\footnotesize
\caption{\footnotesize \label{resultsrealtest} {\it Speech quality through \emph{SRMR} results for real reverberated speech samples.}}
\centerline{
\begin{tabular}{c||c|c|c|c}
Methods & REV-Dev & REV-Eval & VH-v0.2 & VH-v1.0 \\  
\specialrule{.2em}{.1em}{.1em} 
Unprocessed & 3.79 & 3.18 & 3.19 & 4.51 \\ \hline  WPE \cite{NaraWPE} & 4.17 & 3.48 & 3.28 & 4.96 \\ \hline 
CCRN & 4.70 & 4.11 & 5.14 & 5.98 \\ 
+Prog Sup & \cellcolor{gray!35}5.01 & \cellcolor{gray!35}4.44 & \cellcolor{gray!35}6.13 & \cellcolor{gray!35}7.01 \\ \hline 
CCRN-State & 4.65 & 3.87 & 5.09 & 6.43 \\   
+Prog Sup & 4.88 & 4.20 & 5.35 & 0.62 \\ 
\end{tabular}
}
\end{table}
% \vspace{-0.5cm}

%-------------------------------------------------------------------------------------
\subsubsection{Reverberation level, Room sizes, and Near \& Far field}
%-------------------------------------------------------------------------------------
% Figure \ref{fig:srmr}-left shows the evolution of \emph{SRMR} results with the increase of reverberation level for different room sizes: $Room1$, $Room2$, and $Room3$ with $RT_{60}(s) = [0.25,0.5,0.7]$ correspondingly. 

\begin{figure}[h!]
\footnotesize
%\centerline{\epsfig{figure=srmr_full.pdf,width=80mm}}
\centerline{\includegraphics[width=0.9\linewidth]{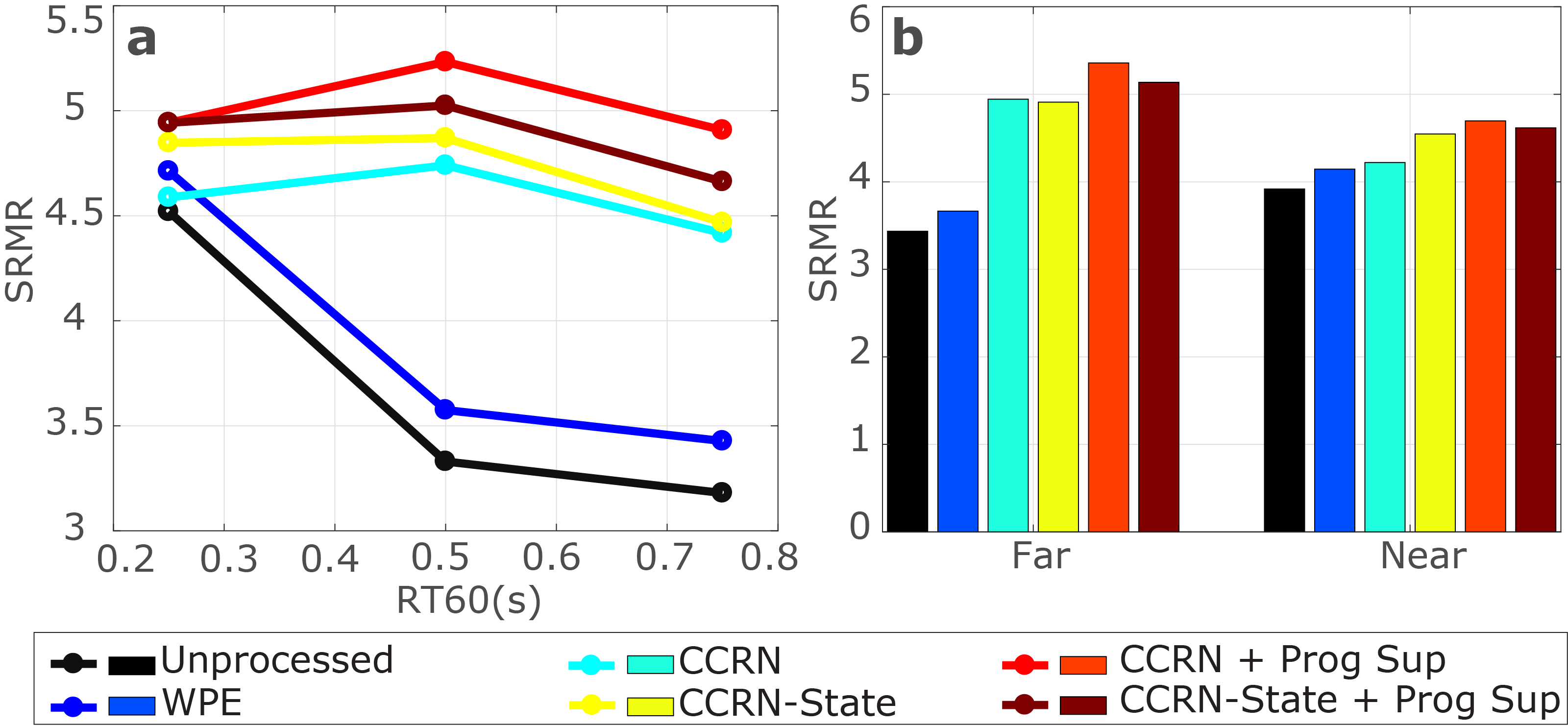}}
\caption{{\it Speech quality through SRMR measure in simulated reverberated speech samples from REVERB Dev \& Eval datasets.}}
\label{fig:srmr}
\end{figure}

Figure \ref{fig:srmr}a shows the evolution of \emph{SRMR} results with the increase of reverberation level for different room sizes and $RT_{60}(s)$. 

%\vspace{-0.5cm}

% All \emph{CCRN}-based methods outperform \emph{WPE} baseline, and consistent with the previous result, \emph{CCRN + Progressive Supervision} achieves the higher speech quality for all conditions evaluated.
All \emph{CCRN}-based methods outperform \emph{WPE} baseline, but \emph{CCRN + Progressive Supervision} achieves the higher speech quality for all conditions evaluated.
Similar behaviour is obtained for $far$ ($250 m$) and $near$ ($50 m$) conditions (Figure \ref{fig:srmr}b).
Note that again the \emph{CCRN + Progressive Supervision} model achieves the best performance, mainly in far-field conditions. 
The introduction of the \emph{Progressive Supervision} stimulates the correct performance of the enhancement process beyond the blind modification of frequency channels performed by \emph{CCRN} and \emph{CCRN-State}. 
It contributes to regularize the network parameters and also provides a progressive transformation of the spectrum towards the final enhancement. 

%We consider that since the spectrum is a time-frequency representation, we think that keeps the relation among the neighbouring areas provide the network with a reference about areas more likely to be corrupted. 
%-------------------------------------------------------------------------------------
% \subsection{Architecture comparison}
\subsection{Architecture discussion}
%-------------------------------------------------------------------------------------

During the enhancement process, architectures \emph{CCRN} and \emph{CCRN-State} provide a messy spectrum that does not supply clues of what the network is doing.
This representation might be a codification of the input, or even it could be learning the training examples.
With \emph{Progressive Supervision}, the networks are able to show the evolution of the predicted signal. 
Moreover, this cost function regularizes the network because it prevents to learn concrete examples.
Finally, we can see how the signal is enhanced at each block. 
We can view the progressive enhancement through residual connections as a spectral power subtraction method. Each RB computes the weighted spectral power, while the residual connection adds it to the corrupted signal.
In this context, the \emph{CCRN + Progressive Supervision} method is comparable to the \emph{WPE} method.
Both, estimate the spectral power to make a subtraction.
\emph{CCRN + Progressive Supervision} uses convolutional layers at each of the \emph{RB}, and \emph{WPE} uses an \emph{LSTM}. 
However, unlike \emph{WPE}, \emph{CCRN + Progressive Supervision} applies as many spectrum subtractions as blocks have the architecture. 
An additional advantage of the \emph{CCRN + Progressive Supervision} is that during training we have access to the reconstruction error at each block.
This allows us to train a big network with many \emph{RB} and then only use the number of \emph{RB} that actually provides significant cost reduction. 
This is a desirable quality when dealing with a clean signal.

%%%%%%%%%%%%%%%%%%%%%%%%%%%%%%%%%%%%%%%%%%%%%%%%%%%%%%%%%%%%%%%%%%%%%%%%%%%%%%%%%%%%%%
\section{Conclusions and Future}
\label{sec:conc}
%%%%%%%%%%%%%%%%%%%%%%%%%%%%%%%%%%%%%%%%%%%%%%%%%%%%%%%%%%%%%%%%%%%%%%%%%%%%%%%%%%%%%%
This paper proposed a deep learning solution for Speech Enhancement based on residual networks. 
By means of an interpretative study of the progressive transformations performed by the network, we were able to design an improved architecture for the speech enhancement purpose.
The mechanism of \emph{Progressive Supervision} contributed to regularize the network parameters. 
It demonstrated to be able to stimulate the correct performance of the enhancement process, beyond the blind modification of frequency channels performed by the other alternative evaluated (\emph{CCRN} and \emph{CCRN-State}). 
The proposal obtained speech enhancement results beyond the state-of-the-art, achieving a favorable trade-off between dereverberation and the amount of spectral distortion. 

We showed that the use of interpretative analysis of the process inside the networks can provide useful insights to develop improved solutions. 
Future work will use the experience at introducing the reconstruction error in each network block for exploring a novel design that prevents an over-enhancement effect.
We plan to define an architecture with many \emph{RB} at training, and during evaluation, to adjust the network size considering the information provided by the reconstruction error. 
%

%%%%%%%%%%%%%%%%%%%%%%%%%%%%%%%%%%%%%%%%%%%%%%%%%%%%%%%%%%%%%%%%%%%%%%%%%%%%%%%%%%%%%%
\section{Acknowledgment}
Funded by the Government of Aragon (Reference Group T36\_17R) and co-financed with Feder 2014-2020 "Building Europe from Aragon". We thanks the Spanish Ministry of Economy and Competitiveness through the project TIN2017-85854-C4-1-R, and NVIDIA Corporation for the donation of the Titan Xp GPU. This material is based upon work supported by Google Cloud.
\bibliographystyle{IEEEtran}
\bibliography{biblio.bib}

\begin{thebibliography}{10}
\providecommand{\url}[1]{#1}
\csname url@samestyle\endcsname
\providecommand{\newblock}{\relax}
\providecommand{\bibinfo}[2]{#2}
\providecommand{\BIBentrySTDinterwordspacing}{\spaceskip=0pt\relax}
\providecommand{\BIBentryALTinterwordstretchfactor}{4}
\providecommand{\BIBentryALTinterwordspacing}{\spaceskip=\fontdimen2\font plus
\BIBentryALTinterwordstretchfactor\fontdimen3\font minus
  \fontdimen4\font\relax}
\providecommand{\BIBforeignlanguage}[2]{{%
\expandafter\ifx\csname l@#1\endcsname\relax
\typeout{** WARNING: IEEEtran.bst: No hyphenation pattern has been}%
\typeout{** loaded for the language `#1'. Using the pattern for}%
\typeout{** the default language instead.}%
\else
\language=\csname l@#1\endcsname
\fi
#2}}
\providecommand{\BIBdecl}{\relax}
\BIBdecl

\bibitem{Xia2013}
B.-Y. Xia and C.-C. Bao, ``Speech enhancement with weighted denoising
  auto-encoder,'' in \emph{Interspeech}, 2013.

\bibitem{Feng2014}
X.~Feng, Y.~Zhang, and J.~Glass, ``Speech feature denoising and dereverberation
  via deep autoencoders for noisy reverberant speech recognition,'' in
  \emph{IEEE International Conference on Acoustic, Speech and Signal Processing
  (ICASSP)}, 2014.

\bibitem{Tu2017}
M.~Tu and X.~Zhang, ``{S}peech enhancement based on deep neural networks with
  skip connections,'' in \emph{IEEE International Conference on Acoustic,
  Speech and Signal Processing (ICASSP)}, 2017, pp. 5565--5569.

\bibitem{Karjol2018}
P.~Karjol, A.~Kumar, and P.~K. Ghosh, ``{S}peech enhancement using multiple
  deep neural networks,'' in \emph{IEEE International Conference on Acoustic,
  Speech and Signal Processing (ICASSP)}, 2018, pp. 5049--5053.

\bibitem{IRASL}
\BIBentryALTinterwordspacing
``{IRASL workshop on behalf of NIPS 2018: Interpretability and Robustness in
  Audio, Speech, and Language}.'' [Online]. Available:
  \url{https://irasl.gitlab.io/}
\BIBentrySTDinterwordspacing

\bibitem{IS2018}
\BIBentryALTinterwordspacing
``{Special Session at Interspeech 2018 ``Deep Neural Networks: how can we
  interpret what they learned?''}.'' [Online]. Available:
  \url{http://interspeech2018.org/program-special-sessions.html}
\BIBentrySTDinterwordspacing

\bibitem{He2016}
K.~He, X.~Zhang, S.~Ren, and J.~Sun, ``Deep residual learning for image
  recognition,'' in \emph{The IEEE Conference on Computer Vision and Pattern
  Recognition (CVPR)}, June 2016.

\bibitem{Qian2017}
K.~Qian, Y.~Zhang, S.~Chang, X.~Yang, D.~Florencio, and M.~Hasegawa-Johnson,
  ``Speech enhancement using bayesian wavenet,'' in \emph{Interspeech}, 2017,
  pp. 2013--2017.

\bibitem{Park2017}
S.~R. Park and J.~Lee, ``{A fully convolutional neural network for speech
  enhancement},'' in \emph{Interspeech}, 2017, pp. 1993--1997.

\bibitem{Santos2018}
J.~F. Santos and T.~H. Falk, ``Speech dereverberation with context-aware
  recurrent neural networks,'' \emph{IEEE/ACM Transactions on Audio, Speech,
  and Language Processing}, vol.~26, no.~7, pp. 1236--1246, 2018.

\bibitem{Pascual2017}
S.~Pascual, A.~Bonafonte, and J.~Serr, ``{Segan: Speech enhancement generative
  adversarial network},'' in \emph{Interspeech}, 2017, pp. 3642--3646.

\bibitem{Santos2018a}
J.~F. Santos and T.~H. Falk, ``Investigating the effect of residual and highway
  connections in speech enhancement models,'' in \emph{Conference on Neural
  Information Processing Systems (NIPS)}, 2018.

\bibitem{Ioffe15}
\BIBentryALTinterwordspacing
S.~Ioffe and C.~Szegedy, ``Batch normalization: Accelerating deep network
  training by reducing internal covariate shift,'' \emph{CoRR}, vol.
  abs/1502.03167, 2015. [Online]. Available:
  \url{http://arxiv.org/abs/1502.03167}
\BIBentrySTDinterwordspacing

\bibitem{He2015}
K.~He, X.~Zhang, S.~Ren, and J.~Sun, ``Delving deep into rectifiers: Surpassing
  human-level performance on imagenet classification,'' in \emph{Proceedings of
  the IEEE international conference on computer vision}, 2015, pp. 1026--1034.

\bibitem{llombart2018wide}
J.~Llombart, A.~Miguel, A.~Ortega, and E.~Lleida, ``Wide residual networks 1d
  for automatic text punctuation,'' \emph{IberSPEECH 2018}, pp. 296--300, 2018.

\bibitem{Zagoruyko16}
\BIBentryALTinterwordspacing
S.~Zagoruyko and N.~Komodakis, ``Wide residual networks,'' \emph{CoRR}, vol.
  abs/1605.07146, 2016. [Online]. Available:
  \url{http://arxiv.org/abs/1605.07146}
\BIBentrySTDinterwordspacing

\bibitem{lee2015deeply}
C.-Y. Lee, S.~Xie, P.~Gallagher, Z.~Zhang, and Z.~Tu, ``Deeply-supervised
  nets,'' in \emph{Artificial Intelligence and Statistics}, 2015, pp. 562--570.

\bibitem{paszke2017automatic}
A.~Paszke, S.~Gross, S.~Chintala, G.~Chanan, E.~Yang, Z.~DeVito, Z.~Lin,
  A.~Desmaison, L.~Antiga, and A.~Lerer, ``Automatic differentiation in
  pytorch,'' in \emph{NIPS-W}, 2017.

\bibitem{garofolo1993darpa}
J.~S. Garofolo, L.~F. Lamel, W.~M. Fisher, J.~G. Fiscus, and D.~S. Pallett,
  ``{DARPA TIMIT acoustic-phonetic continous speech corpus CD-ROM. NIST speech
  disc 1-1.1},'' \emph{NASA STI/Recon technical report n}, vol.~93, 1993.

\bibitem{panayotov2015librispeech}
V.~Panayotov, G.~Chen, D.~Povey, and S.~Khudanpur, ``{Librispeech: an ASR
  corpus based on public domain audio books},'' in \emph{2015 IEEE
  International Conference on Acoustics, Speech and Signal Processing
  (ICASSP)}.\hskip 1em plus 0.5em minus 0.4em\relax IEEE, 2015, pp. 5206--5210.

\bibitem{rousseau2014enhancing}
A.~Rousseau, P.~Del{\'e}glise, and Y.~Esteve, ``{Enhancing the TED-LIUM Corpus
  with Selected Data for Language Modeling and More TED Talks.}'' in
  \emph{LREC}, 2014, pp. 3935--3939.

\bibitem{Kingma2015}
D.~P. Kingma and J.~L. Ba, ``Adam: Amethod for stochastic optimization,'' in
  \emph{Proceedings of the 3rd International Conference on Learning
  Representations (ICLR)}, 2015.

\bibitem{Loshchilov2017}
I.~Loshchilov and F.~Hutter, ``Fixing weight decay regularization in adam,''
  \emph{arXiv preprint arXiv:1711.05101}, 2017.

\bibitem{Challenge2013}
K.~Kinoshita, M.~Delcroix, T.~Yoshioka, T.~Nakatani, E.~Habets, R.~Haeb-Umbach,
  V.~Leutnant, A.~Sehr, W.~Kellermann, R.~Maas, S.~Gannot, and B.~Raj, ``The
  {REVERB C}hallenge: {A} common evaluation framework for dereverberation and
  recognition of reverberant speech,'' in \emph{Proceedings of the IEEE
  Workshop on Applications of Signal Processing to Audio and Acoustics
  (WASPAA-13)}, 2013.

\bibitem{WSJCAMO}
T.~Robinson, J.~Fransen, D.~Pye, J.~Foote, and S.~Renals, ``{WSJCAMO: a British
  English speech corpus for large vocabulary continuous speech recognition},''
  in \emph{IEEE International Conference on Acoustic, Speech and Signal
  Processing (ICASSP)}, 1995, pp. 81--84.

\bibitem{MC-WSJ-AV}
M.~Lincoln, I.~McCowan, J.~Vepa, and H.~Maganti, ``{The multi-channel Wall
  Street Journal audio visual corpus (MC-WSJ-AV): specification and initial
  experiments},'' in \emph{Proceedings of the 2005 IEEE Workshop on Automatic
  Speech Recognition and Understanding (ASRU-05)}, 2005, pp. 357--362.

\bibitem{VoiceHome1}
N.~Bertin, E.~Camberlein, E.~Vincent, R.~Lebarbenchon, S.~Peillon, Éric
  Lamandé, S.~Sivasankaran, F.~Bimbot, I.~Illina, A.~Tom, S.~Fleury, and Éric
  Jamet, ``A french corpus for distant-microphone speech processing in real
  homes,'' in \emph{Interspeech}, 2016.

\bibitem{VoiceHome2}
N.~Bertin, E.~Camberlein, R.~Lebarbenchon, E.~Vincent, S.~Sivasankaran,
  I.~Illina, and F.~Bimbot, ``{VoiceHome-2, an extended corpus for multichannel
  speech processing in real homes},'' \emph{Speech Communications}, vol. 106,
  pp. 68 -- 78, 2019.

\bibitem{NaraWPE}
L.~Drude, J.~Heymann, C.~Boeddeker, and R.~Haeb-Umbach, ``{NARA-WPE: A Python
  package for weighted prediction error dereverberation in Numpy and Tensorflow
  for online and offline processing},'' in \emph{13. ITG Fachtagung
  Sprachkommunikation (ITG 2018)}, Oct 2018.

\bibitem{Kinoshita2017}
K.~Kinoshita, M.~Delcroix, H.~Kwon, T.~Mori, and T.~Nakatani, ``{Neural
  network-based spectrum estimation for online WPE dereverberation},'' in
  \emph{Interspeech}, 2017, pp. 384--388.

\bibitem{Loizou2011}
P.~C. Loizou, \emph{Speech Quality Asssessment. In: Multimedia Analysis,
  Processing and Communications}.\hskip 1em plus 0.5em minus 0.4em\relax
  Springer, 2011, pp. 623--654.

\bibitem{SRMRmeasure}
T.~H. Falk, C.~Zheng, and W.-Y. Chan, ``A non-intrusive quality and
  intelligibility measure of reverberant and dereverberated speech,''
  \emph{IEEE Transaction in Audio, Speech and Language Processing}, vol.~18,
  no.~7, pp. 1766--1774, 2010.

\end{thebibliography}
%%%%%%%%%%%%%%%%%%%%%%%%%%%%%%%%%%%%%%%%%%%%%%%%%%%%%%%%%%%%%%%%%%%%%%%%%%%%%%%%%%%%%%
\end{document}